\documentclass[preprintnumbers,superscriptaddress,pra,showkeys]{revtex4}
\usepackage{amsmath}

\begin{document}

\title{Singular zero-temperature system}
\author{Q. H. Liu}
\email{quanhuiliu@gmail.com}
\affiliation{School for Theoretical Physics, School of Physics and Electronics, Hunan
University, Changsha 410082, China}
\author{S. F. Xiao}
\email{xiaosf@lingnan.edu.cn}
\affiliation{College of Physical Science and Technology, Lingnan Normal University,
Zhanjiang 524048, China}
\affiliation{School for Theoretical Physics, School of Physics and Electronics, Hunan
University, Changsha 410082, China}
\author{Di Guo}
\email{diguo@tjrac.edu.cn}
\affiliation{College of Physics, Tianjin Renai College, Tianjing 524048, China}
\affiliation{School for Theoretical Physics, School of Physics and Electronics, Hunan
University, Changsha 410082, China}
\author{K. J. Yin}
\email{Nitrogennn@hnu.edu.cn}
\affiliation{School for Theoretical Physics, School of Physics and Electronics, Hunan
University, Changsha 410082, China}
\date{\today }

\begin{abstract}
It has long been taken for granted that there is only one type of
thermodynamic system near absolute zero temperature: the ordinary one
compatible with all statements of the third law, with a fundamental yet
tacit assumption that all heat capacities in the system vanish as absolute
temperature approaches zero. However, in the strict sense, the statements
are not mutually equivalent. Once the tacit assumption is released, the
inequivalence must remain, and we may have some systems that are only
compatible with one or two statements but not all, defining a singular
zero-temperature system which can never be excluded from physical
feasibility. We revisit some previously proposed theoretical models and
identify that they belong to the singular system.
\end{abstract}

\keywords{Thermodynamics, Third law, heat capacity, zero temperature,
singular system.}
\maketitle

\section{Introduction}

The third law of thermodynamics imposes the boundary conditions at $T=0$ on
thermodynamics, which is a collection of three inequivalent statements: \cite%
{Landsberg,comment} Nernst statement, Planck statement and Unattainability
statement. By Nernst statement, we mean that the $T=0$ isotherm is also an
isentrope (or "adiabat"), \cite{callen} 
\begin{equation}
\underset{T\rightarrow 0}{\lim }\left( \Delta S\right) _{T}=0,  \label{ns}
\end{equation}%
where we define the constant-temperature entropy change $\left( \Delta
S\right) _{T}$, with a set of independent external variables $\left\{
x\right\} $, 
\begin{equation}
\left( \Delta S\right) _{T}\equiv S\left( T,\left\{ x_{f}\right\} \right)
-S\left( T,\left\{ x_{i}\right\} \right)  \label{endiff}
\end{equation}%
for two equilibrium states $(T,\left\{ x_{i}\right\} )$ and $\left(
T,\left\{ x_{f}\right\} \right) $ $\left( \left\{ x_{f}\right\} \neq \left\{
x_{i}\right\} \right) $ which are connected by a thermodynamic process from
initial values $\left\{ x_{i}\right\} $ to final one $\left\{ x_{f}\right\} $
under constant temperature $T$. By Planck statement, we mean that the
entropy of any pure substance is zero at absolute zero, \cite{comment} i.e., 
\begin{equation}
S\left( 0,V\right) =0,  \label{ps}
\end{equation}%
which is sometimes called as stronger version of the Nernst statement. In
the Nernst statement (\ref{ns}), the temperature $T$ is made to approach
absolute zero but does not necessarily have to be zero, while in the Planck
statement, the zero-point entropy is presupposed to exist. Once $S\left(
0,V\right) $ exists, we have an extremely simple argument to prove why and
how (\ref{ps}) holds true in general. For two different volumes, Nernst
statement implies $S\left( 0,V_{1}\right) =S\left( 0,V_{2}\right) $, i.e., $%
V_{1}s\left( 0,V_{1}\right) =V_{2}s\left( 0,V_{2}\right) $ with $s\left(
0,V\right) $ being entropy density per volume $s\left( 0,V\right) \equiv
S\left( 0,V\right) /V$. Since the entropy density is the same while $V_{1}$
and $V_{2}$ are arbitrarily, we thus must have (\ref{ps}). From this manner
of proof, we see that there is a special case $S\left( 0,V\right) =Ns$ with $%
s\left( 0,V_{1}/N_{1}\right) =s\left( 0,V_{2}/N_{2}\right) =const.$Thus
Planck statement permits a special case%
\begin{equation}
S\left( 0,V/N\right) =const.\neq 0.  \label{sps}
\end{equation}%
We refer to either (\ref{ps}) or (\ref{sps}) as the Planck statement. By the
Unattainability statement, we mean that it is impossible by means of any
process, no matter how idealized, to reduce the temperature of a system to
the absolute zero, in a finite number of steps. \cite{comment} As the
inequivalence of these statements we refer to Reiss, \cite{comment} Callen 
\cite{callen} and also Landsberg (who notes that it has not proved possible
to deduce the unattainability of the absolute zero from the Nernst
statement, since the universe of possible operations of lowering
temperatures can never be fully known.) \cite{Landsberg} For our purposes,
it is beneficial to take these statements as independent statements that
serve different aspects of a system at extremely low temperatures, and by an
ordinary zero-temperature system, we mean that it obeys all statements.

For a second order ordinary differential equation, we usually have two types
of solution, one is analytical in whole domain of the variables and another
is singular at some points, and both solutions are meaningful. Since entropy 
$S$ is not directly measurable, there is no principle to exclude the
possibility of finiteness of $S(0)$ even $S(T=0)=\infty $. \cite%
{comment1,comment2} A system with $S(T=0)\neq 0$, partially compatible with
the third law and fully permissible under other laws, is coined a \emph{%
singular zero-temperature system}. In both mathematics and physics, we have
well-established manners to deal with the singular situations, e.g. making a
singular integral finite, rendering a singular solution of a differential
equation meaningful and renormalizating a divergent series to be convergent,
etc.. Similarly, the singular zero-temperature system can in principle be
physically feasible.

It is seldom emphasized, even mentioned (probably only once explicitly
stated in a note in the literature \cite{comment}) that there is a separate
postulate underlying the current formulation of the third law of
thermodynamics: all heat capacity for a thermodynamic system is zero when $%
T=0$ (all-heat-double-zero (AHDZ), here H, DZ are abbreviated of \textit{h}%
eat capacity, and \textit{d}ouble \textit{z}eros in both heat capacity and
temperature, respectively). In fact, AHDZ can be taken as a statement of the
experimental fact, but far from qualifiable to express a fundamental law for
it is neither sufficient nor necessary for the third law to be held, as we
will discuss in section II and III.

The key purpose of the present study is the introduction of the singular
zero-temperature system and its simple criteria, as well as its
illustrations, which will be done in sections II and III, respectively.
Section IV is the discussion and conclusion.

The present study is confined ourself to thermodynamics without care for the
statistical mechanics. For sake of simplicity, we consider a simple closed,
one-component, one-phase system and choose temperature, $T$, and volume, $V$%
, as our independent variables, and the general form of the equation of
state is $p=p(T,V)$, in which the number of particles $N$ is a parameter
with a fixed value. Our treatment is easily generalized to the complicated
systems.

\section{The singular zero-temperature system: Criteria}

First of all, it is stressed that the precise meaning of AHDZ in standard
discussion of third law is in what follows. The entropy $S\left( T,V\right) $
determined from the constant-volume heat capacity $C_{V}(V,T)$ is given by, 
\begin{equation}
S\left( T,V\right) -S\left( 0,V\right) =\int_{0}^{T}\frac{C_{V}(T^{\prime
},V)}{T^{\prime }}dT^{\prime }.  \label{sx}
\end{equation}%
The ordinary zero-temperature system requires that the entropy at any finite
temperature is to be finite and continuous, which implies both 
\begin{equation}
S\left( 0,V\right) =\text{finite value,}  \label{fund1}
\end{equation}%
and 
\begin{equation}
0\prec \int_{0}^{T}\frac{C_{V}(T^{\prime },V)}{T^{\prime }}dT^{\prime }\prec
\infty \text{ for finite T,}  \label{fund2}
\end{equation}%
simultaneously, rather than $\underset{T\rightarrow 0}{\lim }C_{V}(T,V)=0$
along. Otherwise we have counterexample, the so-called Mattis "electron gas"
system which will be discussed in section III.

Secondly, breakdown of either (\ref{fund1}) or (\ref{fund2}) but agreement
with the consistently modified form of (\ref{ns}) leads to the singular
zero-temperature system. The so called consistently modified form of (\ref%
{ns}) means that once relation (\ref{fund2}) breaks, and we are forced to
deal following singular integral 
\begin{equation}
\underset{\varepsilon \rightarrow 0}{\lim }\int_{\varepsilon }^{T}\frac{%
C_{V}(T^{\prime },V)}{T^{\prime }}dT^{\prime }\rightarrow \infty \text{. }
\label{gi}
\end{equation}%
Thus relation (\ref{sx}) must be modified into following one with $%
\varepsilon $ being very close to zero, 
\begin{equation}
S\left( T,V\right) -S\left( \varepsilon ,V\right) =\int_{\varepsilon }^{T}%
\frac{C_{V}(T^{\prime },V)}{T^{\prime }}dT^{\prime }.  \label{gr}
\end{equation}%
For guaranteeing that at finite temperature $T$ the entropy $S\left(
T,V\right) \ $is finite, we must have from (\ref{gr})%
\begin{equation}
\underset{\varepsilon \rightarrow 0}{\lim }S\left( \varepsilon ,V\right)
\rightarrow -\infty .  \label{zps}
\end{equation}%
The constant-temperature entropy change $\left( \Delta S\right) _{T}$ (\ref%
{endiff}) must be generalized to be 
\begin{equation}
\left( \Delta S\right) _{T}\equiv S\left( T,V_{f}\right) -S\left(
\varepsilon ,V_{f}\right) -\left( S\left( T,V_{i}\right) -S\left(
\varepsilon ,V_{i}\right) \right) .  \label{ds}
\end{equation}%
Special case of the constant-temperature entropy change $\left( \Delta
S\right) _{T}$ is that there is no parameter $\varepsilon $, the definition
of $\left( \Delta S\right) _{T}$ is%
\begin{equation}
\left( \Delta S\right) _{T}\equiv S\left( T,V_{f}\right) -S\left(
T,V_{i}\right)  \label{ds1}
\end{equation}%
The modified Nernst statement is 
\begin{equation}
\underset{\varepsilon \rightarrow 0}{\lim }\underset{T\rightarrow 0}{\lim }%
\left( \Delta S\right) _{T}=0,\text{or}\underset{T\rightarrow 0}{\lim }%
\left( \Delta S\right) _{T}=0  \label{order}
\end{equation}%
where in the first expression we firstly let $T\rightarrow 0$ while $%
\varepsilon $ is given a fixed value and secondly let $\varepsilon
\rightarrow 0$. For an ordinary zero-temperature system, the order of taking
two limits is interchangeable, but for the singular one, it is not.

Lastly, we give the criteria for the singular zero-temperature system in the
following: i) If the system has no external variable, both conditions (\ref%
{zps}) and (\ref{order}) hold true; ii) If the system has a pair of external
variables, e.g., volume and pressure, and for one variable we have
conditions (\ref{fund1}) and (\ref{fund2}) for another variable, we have (%
\ref{zps}) and (\ref{order}); and iii) If the system has more pairs of
external variables, one pair of them must behave similarly as those in case
ii) and other can or cannot do so.

Some comments on the criteria are given in the following.

The integral $\int_{T_{0}}^{T}\frac{C_{V}(T^{\prime },V)}{T^{\prime }}%
dT^{\prime }$ in (\ref{gi}) for a given reference temperature $T_{0}\left(
\prec T\right) $ satisfies $0\prec \int_{T_{0}}^{T}\frac{C_{V}(T^{\prime },V)%
}{T^{\prime }}dT^{\prime }\prec \infty $.

Some prefer to use a specific form of an extensive quantity such as specific
constant-volume heat capacity $c_{V}\equiv C_{V}/N,$and the mean entropy per
particle $s\equiv S/N$, etc. What is more, in cases that $c_{V}\equiv C_{V}/N
$ depends on temperature only, $s\equiv S/N$ is then%
\begin{equation}
s\left( T\right) -s\left( \varepsilon \right) =\int_{\varepsilon }^{T}\frac{%
c_{V}(T^{\prime })}{T^{\prime }}dT^{\prime },  \label{gr1}
\end{equation}%
and modified Nernst statement (\ref{ds}) turns out to be trivially held $%
\underset{\varepsilon \rightarrow 0}{\lim }\underset{T\rightarrow 0}{\lim }%
\left( \Delta s\right) _{T}=0$.

For the singular zero-temperature system, it might be that all heat capacity
for a thermodynamic system is zero when $T=0$. To note that there are two
universal relations which can be easily proved, 
\begin{equation}
\frac{\kappa _{S}}{\kappa _{T}}\equiv \frac{-\frac{1}{V}\left( \frac{%
\partial V}{\partial p}\right) _{S}}{-\frac{1}{V}\left( \frac{\partial V}{%
\partial p}\right) _{T}}=\frac{C_{V}}{C_{p}},\text{and }C_{p}-C_{V}=-T\left( 
\frac{\partial V}{\partial p}\right) _{T}\left( \frac{\partial p}{\partial T}%
\right) _{V}^{2}.  \label{2rs}
\end{equation}%
In consequence, we have three different systems when temperature $T$
approaches to zero, 
\begin{subequations}
\begin{eqnarray}
\text{\textit{A})}\underset{T\rightarrow 0}{\lim }C_{p} &=&\underset{%
T\rightarrow 0}{\lim }C_{V}=\underset{T\rightarrow 0}{\lim }\left( \frac{%
\partial p}{\partial T}\right) _{V}=\underset{T\rightarrow 0}{\lim }\left( 
\frac{\partial V}{\partial T}\right) _{p}=0,  \label{c1} \\
\text{\textit{B})}\underset{T\rightarrow 0}{\lim }C_{V} &=&0\text{, and }%
\underset{T\rightarrow 0}{\lim }C_{p}=-T\left( \frac{\partial V}{\partial p}%
\right) _{T}\left( \frac{\partial p}{\partial T}\right) _{V}^{2}\neq 0,
\label{c2} \\
\text{\textit{C})}\underset{T\rightarrow 0}{\lim }C_{V} &\neq &0\text{, and }%
\underset{T\rightarrow 0}{\lim }C_{p}=C_{V}-T\left( \frac{\partial V}{%
\partial p}\right) _{T}\left( \frac{\partial p}{\partial T}\right)
_{V}^{2}\neq 0.  \label{c3}
\end{eqnarray}%
Once the temperature is finite, all systems are eligible thermodynamic ones
provided they obey $0$th, $1$st and $2$nd law. System \textit{A} is ordinary
one; \textit{B} could be singular (see, the ideal gas in section III); and 
\textit{C} is not applicable to case where the temperature is near to the
absolute zero.

\section{The singular zero-temperature system: Examples}

In literature, we have some systems that are indeed the singular
zero-temperature system, but have not hitherto realized it. Now we revisit
these systems: the Mattis "free electron" system, and Landsberg ideal gas
system, and Zhu system.

\subsection{Mattis "electron gas" system}

Consider a hypothetical system, where the average heat capacity per particle 
$c(T)$ is, \cite{comment2} 
\end{subequations}
\begin{equation}
c(T)=\frac{c_{0}}{\ln (1+T_{0}/T)},  \label{m1}
\end{equation}%
where $c_{0}$ and $T_{0}$ are fixed parameters. One can easily check that $%
c(T)$ goes to zero when $T\rightarrow 0$,%
\begin{equation}
\underset{T\rightarrow 0}{\lim }c(T)=0,  \label{m2}
\end{equation}%
and the entropy $s(0)$ per particle is singular as $s(0)\rightarrow -\infty $%
, as shown by Mattis himself in page of the second edition of Reference \cite%
{comment2}. So, Mattis "electron gas" system violates the Planck statement.
However, the modified Nernst statement (\ref{ds}) holds true for we have $%
\underset{\varepsilon \rightarrow 0}{\lim }\underset{T\rightarrow 0}{\lim }%
\left( \Delta s\right) _{T}=0$.

The original purpose of Mattis constructed such a model is that he hoped to
offer a counterexample of an argument that the vanishing of $c(T)$ at $T=0$
is sometimes (incorrectly) referred to as the third law. \cite{comment2}

\subsection{Landsberg ideal gas system}

The ideal gas can only be jointly defined by equation of state $pV=NkT$ and
internal energy $U=U(T)$ that depends on the temperature only. \cite%
{comment1} We assume that the explicit form of the internal energy is 
\begin{equation}
U=N\frac{\left( kT\right) ^{2}}{4b}.  \label{L1}
\end{equation}%
where $b$ is a constant of dimension of energy. It differs from the
"standard" or classical ideal gas whose internal energy is proportional to
the temperature. The heat capacity is thus 
\begin{equation}
C_{V}=Nk\frac{kT}{2b}\rightarrow 0,\text{when }T=0.  \label{L2}
\end{equation}%
and the entropy can be with $V_{0}$ being a reference volume 
\begin{equation}
S\left( T,V\right) =Nk\ln \frac{V}{V_{0}N}+Nk\frac{kT}{2b}.  \label{L3}
\end{equation}%
The entropy $S\left( T,V\right) $ (\ref{L3}) is compatible with both Planck
and Nernst statements for we have $S\left( 0,V\right) =0$ and $\underset{%
T\rightarrow 0}{\lim }\left( \Delta S(T,V)\right) _{T}=0$.

However, the constant-pressure heat capacity $C_{p}$ is%
\begin{equation}
C_{p}=C_{V}+Nk\rightarrow Nk,\text{when }T=0.  \label{L4}
\end{equation}%
We have also with $\varepsilon $ being very close to zero 
\begin{equation}
S\left( T,p\right) -S\left( \varepsilon ,p\right) =\int_{\varepsilon }^{T}%
\frac{C_{p}}{T^{\prime }}dT^{\prime }=\frac{Nk^{2}}{2b}\left( T-\varepsilon
\right) +Nk\ln \frac{T}{\varepsilon }.  \label{L5}
\end{equation}%
It violates the Planck statement, for we have $S\left( \varepsilon ,p\right)
\sim Nk\ln \varepsilon $ $\rightarrow -\infty $ when $\varepsilon
\rightarrow 0$. The entropy $S\left( T,p\right) $ (\ref{L5}) is independent
from the pressure $p$ thus we have 
\begin{equation}
S\left( T,p_{f}\right) -S\left( \varepsilon ,p_{f}\right) =S\left(
T,p_{i}\right) -S\left( \varepsilon ,p_{i}\right) .  \label{L7}
\end{equation}%
The modified Nernst statement (\ref{ds}) holds true for we have%
\begin{equation}
\left( \Delta S(T,p)\right) _{T}\equiv S\left( T,p_{f}\right) -S\left(
\varepsilon ,p_{f}\right) -\left( S\left( T,p_{i}\right) -S\left(
\varepsilon ,p_{i}\right) \right) =0.  \label{L8}
\end{equation}%
\newline

This ideal gas system is clearly not a classical one. This system was
originally constructed by Landsberg with the same purpose as that of Mattis. 
\cite{comment1}

\subsection{Zhu "solid" system}

The equation of state is given by $p=p(T,v)$ \cite{comment3} 
\begin{equation}
p=\frac{a_{1}T^{2}+a_{2}}{(v-a_{3})^{1/2}}  \label{Z1}
\end{equation}%
where $a_{i}$ ($i=1,2,3$) are three constants. The internal energy $u(T,v)$
and entropy $s(T,v)$ are, respectively, \cite{comment3} 
\begin{subequations}
\begin{eqnarray}
u &=&k(T-T_{0})+2(a_{1}T^{2}-a_{2})(v-a_{3})^{1/2},  \label{Z2} \\
s &=&k\ln \frac{T}{T_{0}}+4a_{1}T(v-a_{3})^{1/2}.  \label{Z3}
\end{eqnarray}%
The specific heat capacities are, respectively, \cite{comment3} 
\end{subequations}
\begin{equation}
c_{v}=k+4a_{1}T(v-a_{3})^{1/2},c_{p}=c_{v}+c(v)T^{3}  \label{Z4}
\end{equation}%
where $c(v)$ is a nonvanishing coefficient depending on specific volume $v$
and $a_{i}$. Both $\underset{T\rightarrow 0}{\lim }c_{p}=\underset{%
T\rightarrow 0}{\lim }c_{v}=k$, and $s\rightarrow -\infty $ as $T\rightarrow
0$ violates the Planck statement, but modified Nernst statement is valid $%
\left( \Delta s\right) _{T}=4a_{1}T\left(
(v_{f}-a_{3})^{1/2}-(v_{i}-a_{3})^{1/2}\right) \rightarrow 0$ as $%
T\rightarrow 0.$

Why Zhu system can be called "solid" is due to two response functions
vanishes as $T\rightarrow 0$,%
\begin{equation}
\left( \frac{\partial p}{\partial T}\right) _{V}=\frac{2a_{1}T}{%
(v-a_{3})^{1/2}},\left( \frac{\partial V}{\partial T}\right) _{p}=\frac{%
4a_{1}(v-a_{3})T}{a_{1}T^{2}+a_{2}}.  \label{Z5}
\end{equation}%
This system originally constructed by Zhu was to give a counterexample to
claim that the derivation of $\underset{T\rightarrow 0}{\lim }c_{p}=\underset%
{T\rightarrow 0}{\lim }c_{v}=0$ from Nernst statement is possible. \cite%
{comment3} However, because his system is a singular system, the modified
Nernst statement (\ref{ds1}) rather than the ordinary one needs to be used.

\section{Discussions and Conclusions}

Near absolute zero temperature, the systems can be classified into two
categories. One is ordinary one and another is the singular one, depending
on whether it compatible with all statements of the third law or with only
the modified Nernst statement. For a singular zero-temperature system, the
Unattainability statement is not applicable. In this sense, the existence of
the singular zero-temperature system zooms in the inequivalence between
different statements of the third law. 

It seems that the existence of the singular zero-temperature system is
relevant to the experimental facts on AHDZ or not. In fact, the precise
condition directly related to the heat capacity is in a discriminant
integral (\ref{fund2}), rather than the heat capacity itself. Mattis system
shows that when a heat capacity for a thermodynamic system is zero when $T=0$
but the system violates the discriminant integral (\ref{fund2}), and the
Planck statement of the third law is not longer valid. 

Some may regard that the ideal gas does not exist near zero temperature.
This misunderstanding is caused by (incorrectly) referring to the classical
ideal gas as the general one. We explicitly demonstrate that in general the
ideal gas is a singular zero-temperature system\emph{\ }when temperature
approaches to zero, for we have volume variable version $\left\{
C_{V}(T,V),S(T,V)\right\} $ that behave like an ordinary system whereas the
pressure variable version $\left\{ C_{p}(T,p),S(T,p)\right\} $ is singular
at $T=0$. 

For a singular system, the AHDZ alone is not a precise criteria to
distinguish the ordinary and the singular system. In this sense, AHDZ can
never be taken as a fundamental postulate. Here, we comment on why we are
against the belief that AHDZ can be taken as a postulate from point of view
of more philosophy of physics. Fundamental quantities in physics, such as
the Lagrangian, Hamiltonian, wave function, electromagnetic potentials,
entropy, and absolute zero temperature, are often experimentally
unmeasurable or inaccessible. \cite{dyson} In thermodynamics, entropy and
absolute temperature are a pair of conjugate quantities introduced
simultaneously, yet both have experimentally unattainable aspects: entropy
cannot be directly measured, and absolute zero ($0$K) is unreachable. Here
we like to quote an observation of Callen: "The question of whether the
state of precisely zero temperature can be realized by any process yet
undiscovered may well be an unphysical question, raising profound problems
of absolute thermal isolation and of infinitely precise temperature
measurability. " \cite{callen} Since heat capacity is directly measurable
quantity, we can conclude that AHDZ can never be taken as a fundamental
postulate.

\begin{acknowledgments}
This work is financially supported by the Hunan Province Education Reform
Project under Grants No. HNJG-2023-0147. QH is indebted to the members of
Online Club Nanothermodynamica (Founded in June 2020), and members of
National Association of Thermodynamics and Statistical Physics Teachers in
China, for fruitful discussions.
\end{acknowledgments}

\end{document}